\def \s{~\rm{s}}
\def \km{~\rm{km}}

\def \K{~\rm{K}}

\def \AU{~\rm{AU}}

\def \yr{~\rm{yr}}

\documentclass[12pt,preprint]{aastex}
\usepackage{natbib}
\usepackage{epsfig}
\begin{document}   

\title{Shaping Planetary Nebulae and Related Objects}

\author{Noam Soker\altaffilmark{1}}
\affil{Department of Physics,
Technion$-$Israel Institute of Technology, Haifa 32000, Israel,
Email: soker@physics.technion.ac.il}

\altaffiltext{1}{Also Department of Physics, Oranim}





\begin{abstract}          
I review some open questions and other aspects concerning the shaping of
planetary nebulae (PNs) and related objects.
I attribute the non-spherical structures of PNs to binary
companions, stellar or substellar.
I emphasize the role of jets (or collimated fast wind: CFW)
blown by an accreting stellar companion in shaping bipolar PNs
and some related objects, and discuss the limited role of
magnetic fields.
I speculate that some stars which are leaving the
asymptotic giant branch, i.e., becoming hotter, possess long-term
(10-1000 years) oscillations; these may be related to semi-periodic
concentric rings.
I end with a list of objects whose shaping is related to
the shaping processes of PNs, from young stellar objects
to clusters of galaxies.
\end{abstract}

\section{Introduction}
This paper addresses some of the following open questions related
to the shaping of planetary nebulae (PNs), with a number of
comparisons to related objects.
\begin{enumerate}
\item Why and when does mass loss rate from asymptotic giant branch
(AGB) progenitors of PNs become non-spherical?
Many AGB stars lose mass in a spherical geometry, while most
PNs are axisymmetric, or point symmetric. At the end of the AGB
the mass loss geometry must change to become non-spherical.
I address this question, and show that the binary model for the
shaping of PNs can account for it, in section 2.
\item What is the connection of multiple semi-periodic concentric
shells (arcs; rings; termed M-arcs) to the previous question?
These are thought to be concentric (more or less) semiperiodic
shells; some shells are complete while others are not.
Reviews of the arc and ring properties are given by
Hrivnak, Kwok, \& Su (2001), Kwok, Su, \& Stoesz (2001),
Su (2004), and Corradi et al.\ (2004).
I address this question in section 4.
\item When does a star in the ``right'' initial mass range
form a PN? I will not address this issue directly here,
but it is important when comparing theory with observations, e.g.,
in population synthesis studies.
Soker \& Rappaport (2000) discuss this issue.
In particular, some stars will not form PNs: either they lose most
or all of their envelope before reaching the upper AGB, or their
mass loss rate while terminating the AGB is too low, and the PN
will be too faint to be detected.
This issue is related to questions raised be De Marco (2004)
on the number of PNs with binary central stars.
\item What are the roles of jets in shaping PNs?
If jets are not well collimated, I term them CFW, for collimated fast
wind.
In this respect, I will present the view that most, or even all,
jets in PNs are blown by a binary companion--or even by a tertiary star.
I elaborate on this question in relation to jets in clusters
of galaxies in section 5.
\item What is the role of axisymmetric mass loss from the
AGB stellar progenitor?
Models for enhanced equatorial mass loss rate include fast rotating
AGB stars, which were spun-up by companions (Soker 1997), and
effects of magnetic fields.  I will mention the presence of
magnetic fields on the surface of AGB stars, and present my view
that even when magnetic fields are detected, they are not likely
to play a global (as opposed to local) dynamical role in the mass
loss process from the AGB star.
In section 3 I will try to clarify some confusion related to the
roles of magnetic fields in shaping PNs.
\end{enumerate}

Although this paper is basically a review following the
Asymmetrical Planetary Nebulae III (APN3) meeting, it contains
some new ideas as well. It doesn't of course address all
relevant questions, and doesn't cite many papers; many
of these can be found
in papers already cited here.
Note in particular the recent review by Balick \& Frank (2002),
which contains many references and relevant discussions, as do many
papers in the Proceedings of APN3 (I hope all will
be posted on astro-ph).

\section{Basic Types of Binary Interaction}

I distinguish between four basic types of interaction with stellar
or substellar companions. Two of these lead to the formation of
axisymmetrical PNs.
\begin{enumerate}
\item No interaction.
There is no companion to the AGB progenitor, or the companion has
too little mass and/or the orbital separation is too large.
In that case a spherical PN is expected, and/or
the mass loss rate stays low, and no observable PN ever emerges.
\item A very-wide companion.
The companion does not influence the mass loss process because the
orbital separation is too large, $a \gtrsim 100 \AU$.
However, the companion is close enough, $a \lesssim 10^4 \AU$,
and massive enough, i.e., a stellar companion, such that it
causes a departure from axisymmetry (see Soker \& Rappaport 2001
for detailed study of this process, and references therein;
Classification of PNs according to their departure from axisymmetry
is reviewed by Bobrowsky 2004).
\item Shaping without disturbing the AGB progenitor.
In this type of interaction the companion deflects the AGB wind,
but it doesn't influence the AGB progenitor itself.
In the main process (Soker  2001a) the compact companion accretes
from the AGB wind such that an accretion disk is formed, resulting
in two jets (or a CFW).
Soker (2001a) finds that $\sim 5-20 \%$ of all PNs are formed by this
process.
\item Shaping by disturbing the AGB progenitor.
This type of interaction includes tidal interaction,
Roche lobe over flow (RLOF), and common envelope.
RLOF and tidal interaction (RLOF implies tidal interaction,
but tidal interaction can occur without RLOF; Soker 2002c)
are likely to lead to the formation of jets, and the formation
of a bipolar PN with a narrow waist between the two lobes
or bubbles (Soker \& Rappaport 2000).
During the common envelope phase jets are not likely
to be blown, but the companion may blow jets prior to entering
the CE or immediately after.
Common envelope models include substellar companions
(planets and brown dwarfs) as well.
\end{enumerate}
The different types of interaction, the nature of the companion,
and more on their relation to the shaping of PNs are in
some of my earlier papers (e.g., Soker 1997; 1998; 2002c;
Soker \& Rappaport 2000).
Note that a very-wide companion can exists together with
a closer companion.
This is indeed the case in many PNs showing departure
from axisymmetry or from point-symmetry (Bobrowsky 2004).

In the binary model for shaping PNs, i.e., in the last two basic
types of interaction, it is expected that the shaping mainly
occurs as the progenitor is about to leave the AGB, and during the
early post-AGB phase (for related phenomenon see the discussion of
subdwarf B binaries in section 6). When the companion is at a
large orbital separation (interaction type 3), the formation of a
stable accretion disk, and presumably the formation of a CFW (or
two jets), requires the AGB wind to be intensive and slow (Soker
2001a). Practically, for a significant fraction of binary systems
to blow jets ($\gtrsim 10 \%$ of initial binary systems), the AGB
mass loss rate should be $\dot M_1 \gtrsim 10^{-6} M_\odot
\yr^{-1}$ for a wind speed of $v_w \sim 7 \km \s^{-1}$, and $\dot
M_1 \gtrsim 10^{-5} M_\odot \yr^{-1}$ for a wind speed of $v_w
\sim 11 \km \s^{-1}$ (Soker 2001a). As mass loss rate increases in
the range $\gtrsim 2 \times 10^{-6} M_\odot \yr^{-1}$, the AGB
wind speed decreases (Elitzur, Ivezic \& Vinkovic 2002).
Therefore, this condition, of very high mass loss rate and a slow
wind, is met only during the so-called superwind phase of the AGB,
when mass loss rate is very high and wind speed is low. Huggins et
al.\ (2003) find the flow structure in the bipolar proto-PN He
3-1475 to support such scenarios, i.e., where enhanced mass loss
by the central star leads to ejection of two jets.

 In the second
basic type of interaction that leads to the formation of
axisymmetrical PNs (type 4 interaction above), the companion
substantially spins-up the AGB envelope, in particular during the
strong tidal interaction phase, prior to the onset of RLOF and
common envelope. Most of the spin-up, i.e., deposition of orbital
angular momentum to the AGB envelope, occurs during a very short
time (Soker 1995). This is likely to enhance mass loss rate.
Therefore, this type of interaction both increases the mass loss
rate and changes it to axisymmetric, rather than spherical. The
high mass loss rate implies that the AGB progenitor is about to
leave the AGB shortly.

One conclusion from the correlation between AGB high mass
loss rate and the transition to axisymmetry, as observed and as
expected in the binary model, is that most
spherical PNs will not show signatures of superwind.
This is indeed the case when one carefully defines spherical
PNs (Soker 2002a, $\S 2.2$).

\section{Comments on the Role of Magnetic Fields}

The previous section, which clarified some aspects of binary
interaction and the shaping of PNs, would not be complete
without clarification of some aspects of magnetic fields.

\subsection{The source of magnetic fields}
There are now clear indications of magnetic fields
in PNs and around AGB stars.
The strongest indication comes from polarized maser emission
(e.g., Kemball \& Diamond 1997; Zijlstra et al.\ 1989;
Miranda et al.\ 2001a; Vlemmings, Diamond, \& van Langevelde \ 2002;
Indra 2004).
In many cases the {\it local} magnetic pressure inferred
from the maser polarization is about equal to, or even larger
than, the thermal pressure of the gas.
Some papers claim that the strong magnetic fields indicate that
magnetic fields globally shape PNs.
I think that this conclusion is hardly justified.
Instead, I (Soker 2002d) argue for magnetic fields with small
coherence lengths, which result from stellar magnetic spots or
from jets blown by an accreting companion;  the magnetic field does
not play a global role in shaping the circumstellar envelope.
In the solar wind, in general, magnetic pressure exceeds thermal
pressure only in magnetic clouds  (e.g., Yurchyshyn et al.\ 2001),
which are formed by impulsive mass loss events from the sun.
In Soker \& Kastner (2003) we suggest that the maser spots with
strong magnetic fields which are observed around some AGB stars
are similar in nature to the magnetic clouds in the solar wind,
in that they represent local enhancements of the magnetic field.
In related papers, Dorch \& Freytag (2002) discuss local dynamo in
Betelgeuse, and in Soker (2002e) I discuss enhanced magnetic activity
of the cool companion in symbiotic systems.

To summarize this subsection, I expect the global magnetic pressure
in the wind of AGB stars to be much below the thermal pressure,
as winds from AGB stars are driven by radiation pressure on dust
rather than by magnetic activity as in the sun.
The magnetic field results from one of the following.
(1) Locally (but not globally) strong magnetic field spots.
(2) A magnetically active main sequence companion.
(3) An accretion disk around a companion, which amplifies
the magnetic field of the accreted gas.

\subsection{The role of magnetic fields}
Models where magnetic fields play a role in shaping the circumstellar
matter are of four kinds.
(1) Magnetic fields deflect the flow close to the stellar surface,
i.e., the magnetic pressure and/or tension are dynamically
important already on the AGB (or post-AGB) stellar surface
(e.g., Pascoli 1997; Matt et al.\ 2000; Blackman et al.\ 2001).
(2) The magnetic field is weak close to the stellar surface.
It plays a dynamical role only at large distances from the AGB star
(e.g., Chevalier \& Luo 1994; Garc\'{\i}a-Segura 1997;
Garc\'{\i}a-Segura et al.\ 1999).
(3) Locally enhanced magnetic fields on the AGB stellar surface
form cool spots.
Dust formation rate, hence mass loss rate, is enhanced above these
cool spots (Soker \& Zoabi 2002, and references therein).
Magnetic fields never become dynamically important on a large scale.
This mechanism can operate even for very slowly rotating
AGB envelopes; spin-up by planets is enough to account for
the required rotation velocity.
This process may lead to the formation of moderate elliptical PNs
(those with small departure from sphericity), but can't
account for lobes, jets, etc.
(4) Magnetic fields play a dominant role in the launching of jets
from accretion disks, either around stellar companions or
around the central star.

My view that the first two processes mentioned above
(for more on these see Garc\'{\i}a-Segura, Lopez, \& Franco 2004;
Matt, Frank, \& Blackman 2004; Blackman 2004)
can't work in shaping PNs was presented
in several papers (e.g., Soker \& Zoabi 2002).
In particular, they require that the progenitor AGB (or post-AGB)
envelope be spun-up by a stellar companion.
The companion then has other effects on the mass loss process
which are more influential.
It is important to emphasize that any model based on magnetic activity
must state the required angular velocity and/or angular momentum,
and then the source of this angular momentum.
Models for magnetic activity during the post-AGB phase,
have in addition the problem of explaining the bipolar outflow
from systems where the progenitor is still an AGB star.

Process (4) is of a different nature.
From young stellar objects, active galactic nuclei, and many other
systems, it is well established that accretion disks can form two
jets.
The question regarding the exact mechanism for launching
the two jets is an open one.
However, for the pure question of PNs shaping, the main issue
is whether an accretion disk is formed or not; understanding the
exact launching mechanism will not solve the major open questions
regarding the shaping of PNs.

In light of the problems mentioned in this subsection
(and which are discussed in more detail in some of my
earlier papers cited above),
I have the feeling that the APN3-meeting talks devoted to magnetic
activity were much too optimistic in estimating
their ability to shape PNs.

\section{Post-AGB Envelope Instabilities}

\subsection{Personal note}
The evolution of a single star during the final AGB phase
(as it is about to leave the AGB) and early
post-AGB phase holds some unsolved puzzles
(even before adding the complications due to a binary companion).
Some of these will be mentioned later in this section.
My view is that present numerical codes can't handle all
these complications, which involve processes on
timescales and lengthscales which span a large domain, e.g.,
pulsations with large amplitudes on the surface, possible
magnetic activity with local dust formation close to the surface,
large convective eddies, long-term envelope relaxation
(see later this section).
In several papers, and in this section, I address some of
these issues.
Two recent papers (Soker 2003c; Soker \& Harpaz 2003) were
rejected by journals following referee reports.
I think the arguments in these papers still hold, even
if containing some speculative components.
I have very strong criticisms on these referee reports,
but I will not present them for the following reason.
My paper on the magnetic activity of the cool component in
symbiotic systems (Soker 2002e) received a very negative referee
report (it was published following a positive report by a second
referee).
I presented the entire report with my strong criticism
at a meeting on symbiotic stars in 2002.
I submitted these for the Proceedings of that meeting, but was
asked by the editors to remove this report and my criticism from
the Proceedings. In the near future, therefore, I will not
repeat such an attempt (it is not a common practice in
our community).
I think that the present situation, where someone can present his
or her scientific view (i.e., in a referee report) without the need
to stand behind this view (because it is not going to be published
anywhere) is scientifically intolerable.
The electronic publication system must be used to change the
refereeing system.
For example, a paper can be posted in a public place for a few
weeks for open comments from anyone interested.
Should someone think that the paper should not be published, his
or her view (with the name) can be given along with the paper, and
with the authors' reply; future readers will make their own
decision in the dispute.

\subsection{Indications for long-term instabilities}
The evelopes of upper AGB stars are known to be unstable, not only
to oscillations on dynamical time scales of $\sim 1 \yr$, but
also to much longer time-scale perturbations.
On the theoretical side there are several effects that
may lead to such long-time scale variations.
Icke, Frank, \& Heske (1992) found chaos in the behavior of
oscillating AGB stars on time-scales much longer than the
dynamical time.
Ya'ari \& Tuchman (1996, 1999) emphasize the need to include
long-term nonlinear thermal effects, which change the entropy
structure of the envelope, when analyzing pulsating AGB stars.
The dependence of the opacity on temperature, in particular
in the temperature range appropriate for the photospheres
of upper AGB stars, may also lead to sharp changes in the
AGB stellar radius (Soker 2003c).
This effect is particularly important in oxygen-rich AGB stars
because the opacity sharply increases as temperature drops below
$\sim 2900 \K$, and it may lead to large expansion, a process termed
opacity-induced over-expansion (Soker 2003c).
This may occur, therefore, when the photospheric (effective)
temperature drops to $T_p \sim 2900 \K$.
The much higher opacity implies a much lower photospheric density,
which allows the envelope to expand
(I will return to this effect later in this section).
Finally, when the envelope mass becomes low the AGB star stops
expanding and starts contracting (see below).
The numerical stellar evolutionary model kindly given to me
by Amos Harpaz (Soker 1992) shows the stellar envelope
shrinking a little when the envelope mass becomes
$M_{\rm env} \simeq 0.16 M_\odot$, and then expanding again
when $M_{\rm env} \simeq 0.08 M_\odot$.

From the observational direction, Zijlstra, Bedding, \& Mattei
(2002) studied the long-term period evolution of the Mira variable
R Hydrae.
They find the orbital period to decrease from 495 days to 385 days
in $\sim 200 \yr$, from which they deduce
a decrease of the stellar radius by $\sim 16 \%$.
They also argue for a strong decline in the mass loss rate,
by a factor of $\sim 10$, with the envelope contraction.
Zijlstra et al.\ (2002) then very nicely raise the possibility
that if the behavior of R Hydrae is periodic, it can account
for multiple semi-periodic concentric shells (arcs; rings;
I term them `M-arcs').

\subsection{Motivation for long-term semi-periodic evolution}
As indicated in the previous subsection, there are strong
arguments in favor of long-term semi-periodic variations
in upper-AGB stars. I now show that such behavior may resolve
some puzzling issues.
\newline
{\bf (1) M-arcs.} Zijlstra et al.\ (2002) already proposed that
long-term non-linear behavior like that in R Hydrae can form
M-arcs.
Presently there is a disagreement on the process that forms
M-arcs. In the literature there are several proposed models:
($i$) Binary interaction. It seems that this model
can't work (see Soker 2002b). ($ii$) Instability in dust-disk
coupling (Deguchi 1997; Simis, Icke, \& Dominik 2001;
Meijerink, Mellema, \& Simis 2003).
Although this model is popular, I think it has some problems
(Soker 2002b).
($iii$) Solar-like magnetic activity cycle (Soker 2000;
Garc\'{\i}a-Segura, Lopez, \& Franco 2001).
As more PNs are found to have M-arcs (Corradi et al.\ 2004),
this speculative idea becomes less likely.
($iv$) Van Horn et al.\ (2003) propose that the influence of mass
loss on the H-burning shell can set long-term semi-periodicity
in the nuclear burning rate and mass loss rate.
A crucial assumption in their model is that the mass flux
throughout the envelope is constant and equals the mass
loss flux at the stellar surface. I consider this assumption
unrealistic, as numerical models show the envelope mass
decreasing (e.g., Soker 1992; Soker \& Harpaz
1999\footnote{Note that the density scale in Figs.\ 1-5 of
Soker \& Harpaz (1999) is too low by a factor of 10;
the correct scale is displayed in their Fig. 6.}),
rather than mass being supplied to the envelope by the core.
It seems the nuclear burning rate will not
be influenced by the mass loss process.
($v$) Long-term envelope instabilities (Zijlstra et al.\ 2002).
In the next subsection I will propose a phenomenological
model for such a behavior.
In any case, it seems that the mass-losing star blows the
M-arcs before being disturbed. Therefore, I speculate that most
PNs with M-arcs come from binary interation type 3 of section 2,
i.e., the companion does not disturbed much the AGB star, and shaping
occurs mainly via accretion and jet blown by the companion.
Type-4 interacting binaries can also lead to M-arcs if the interaction
occurs at a late stage.
The images of most PNs with M-arcs (Corradi et al.\ 2004) look
like spherical shells that were shaped by jets, and most
don't have large bubbles and narrow waist. Large bubbles and narrow
waist are likely to reuslt from strong interaction (Soker \& Rappaport
2000) that may prevent spherical M-arcs

\noindent {\bf (2) Mass loss rate and transition time.}
AGB stars start to shrink and their effective temperature
starts to increase when their envelope mass decreases to
$M_{\rm env} \simeq 0.2-0.8 M_\odot$, with the higher values for more
massive cores.
In many models for dust formation (e.g., Wachter et al.\ 2002),
the mass loss rate steeply decreases as the temperature increases.
The strong dependence on the temperature implies that as the star
evolves along the post-AGB track and becomes hotter, at constant
luminosity, the mass loss rate steeply decreases with time.
However, observations of planetary nebulae (PNs) and
stellar evolution calculations along the post-AGB require the
high mass loss rate to continue during the early post-AGB phase
(e.g., Tylenda \& Stasinska 1994).
Models where the stellar effective temperature is the sole main
physical parameter which determines the mass loss rate, therefore,
predict too long post-AGB evolution
(for more on this see Soker \& Harpaz 2003).
Soker \& Harpaz (2003) argue that the envelope structure, in particular
the entropy and density gradients, should be among the main
parameters which determine the mass loss rate on the tip of
the AGB and the early post-AGB evolutionary phases.
The entropy profile becomes steeper and the density profile becomes
shallower as the star becomes hotter on the early post-AGB phase,
until the star heats up to $T \gtrsim 8000\K$.
Soker \& Harpaz (2003) propose that mass loss rate stays very high
because of the envelope structure, and drops only when the effects
of the temperature become important once again as the post-AGB
star heats up to $\sim 6,000 \K$.
Note, they do not propose a new mass loss mechanism, but rather mention
several mechanisms by which these profiles may influence the mass
loss rate within the popular mechanism for mass loss on the AGB,
where pulsations coupled with radiation pressure on dust
cause the high mass loss rate
(e.g., H\"ofner 1999; Winters et al.\ 2000;
Andersen, H\"ofner \& Gautschy-Loidl 2003; Willson 2004).
This discussion suggests that some effect(s) in addition to the
dependence on effective temperature (or stellar radius and luminosity)
must operate during the upper AGB phase.
Long-term semi-periodic oscillations, resulting from the steep
entropy profile and shallow density profile, are discussed
in the next subsection.

\subsection{Phenomenological treatment of long-term oscillations}
When the envelope mass of an AGB star declines to
$\lesssim 0.2 M_\odot$ the density profile becomes very shallow
and the entropy profile very steep as the envelope mass decreases
(see graphs in Soker 1992, and Soker \& Harpaz 1999).
This can be seen from simple principles (Soker \& Harpaz 1999, 2003;
Soker 2003c, where the quantitative derivation is given).
The photospheric density, derived from the definition of
the optical depth being $2/3$, is inversely proportional to the
the photospheric temperature $T_p$, stellar radius $R$, and
opacity $\kappa$: $\rho_p \propto (R^2 \kappa T_p)^{-1}$.
As the star evolves along the AGB, the temperature drops and
the opacity drops as well, increasing the photospheric density.
Because of mass loss and increasing radius at the same time, the
average density in the envelope decreases.
The outcome is that the density profile in the envelope becomes
very shallow.
To maintain the convective energy flux in a low density envelope,
the entropy profile becomes very steep (Soker \& Harpaz 1999).
The outcome is an envelope prone to instabilities.
Since the photospheric density must be lower than the average
envelope density, the envelope must eventually shrink to increase
the average envelope density and lower the photospheric density.
This is the reason AGB stars start to shrink and heat up when
their envelope still contains substantial mass.
For a core mass of $M_c \simeq 0.6 M_\odot$ the AGB star
starts shrinking when $M_{\rm env} \simeq 0.2 M_\odot$
(for $M_c \simeq 0.9$ the AGB star starts shrinking earlier,
when $M_{\rm env} \simeq 0.8 M_\odot$; see data in Frankowski 2003).
Using the data given by Frankowski (2003; also private communication),
and Soker (1992), I take a simple formula to describe the
equilibrium, i.e.,  non-oscillatory, average stellar radius as
a function of envelope mass in the range
$0.02 < M_{\rm env} < 0.2 M_\odot$.
I take
\begin{equation}
R_{\rm eq} =260+1400 M_{\rm env}-3500 M^2 _{\rm env},
\end{equation}
where envelope mass is in units of solar mass and radius
in units of solar radius.
This function is drawn as a thick line on the upper panel of
Figure 1.

The shallow density profile has the following implication when
mass loss rate is high.
To maintain a constant stellar radius the lost mass must
be replaced by mass from inner layers.
Because the envelope density profile is shallow, to maintain
a negative density profile, mass must flow outward from deep layers.
This process increases the gravitational energy of
the envelope; the source of this energy is the nuclear burning,
i.e., the stellar luminosity.
Because the envelope stays in a quasi-equilibrium, only a small
fraction, $\eta$, of the stellar luminosity, $L_{\rm eq}$,
will be used to lift mass.
Let me take a high mass loss rate of $\sim 10^{-4} M_\odot \yr^{-1}$,
such that in $\sim 10 \yr$ a mass of $\Delta m \simeq 10^{-3} M_\odot$
is lost.
(For an AGB star with $L_{\rm eq}=6000 L_\odot$ the maximum
mass loss rate will be lower by a factor of 3-5, compared with the
scaling used here.)
I also scale the equation with $\eta = 10^{-3}$,
$L_{\rm eq} =6000 L_\odot$, and a core mass of $M_c = 0.6 M_\odot$.
To lift mass from an average radius of
$r_{\rm in} \sim 0.1 R_{\rm eq} \sim 30 R_\odot$ to large radii,
the thermal time scale is
\begin{equation}
\tau \simeq \frac{G M_c \Delta m}{r_{\rm in} \eta L} \simeq
100 \left( \frac {\Delta m}{0.001 M_\odot} \right)
\left( \frac {r_{\rm in}}{30 R_\odot} \right)^{-1}
\left( \frac {\eta L_{\rm eq}}{6 L_\odot} \right)^{-1}  \yr .
\end{equation}

To summarize the previous paragraph, I propose that the shallow
density profile on the upper AGB and early post-AGB phases
implies that the envelope shrinks after a short period of high
mass loss rate.
After a time-scale of $\sim 100 \yr$, which is basically a thermal
time-scale, the envelope re-expands.
Because of the steep entropy profile the envelope is prone to
instabilities, and it rebounds to larger than the equilibrium radius.
For demonstrative purposes only, I build the following
qualitative phenomenological model.
I take the envelope to possess long-term oscillations, as depicted by the
thin line in the upper panel of figure 1.
The exact form of the assumed long-term oscillation is not
important for the present demonstration.
In any case, the form used is given by
\begin{equation}
R_{long} = R_{\rm eq} +A_r,
\end{equation}
where
\begin{equation}
A_r= - 0.2 R_{\rm eq} \sin^2 [2 \pi(0.2-M_{\rm env})/0.02]
\qquad {\rm for} \qquad
\sin [2 \pi(0.2-M_{\rm env})/0.02] \leq 0
\end{equation}
and
\begin{equation}
A_r= 0.25 (440-R_{\rm eq}) \sin^{1/2} [2 \pi(0.2-M_{\rm env})/0.02]
\qquad {\rm for} \qquad
\sin [2 \pi(0.2-M_{\rm env})/0.02] > 0.
\end{equation}
Masses and radii are in solar units.
\begin{figure}
\centering\epsfig{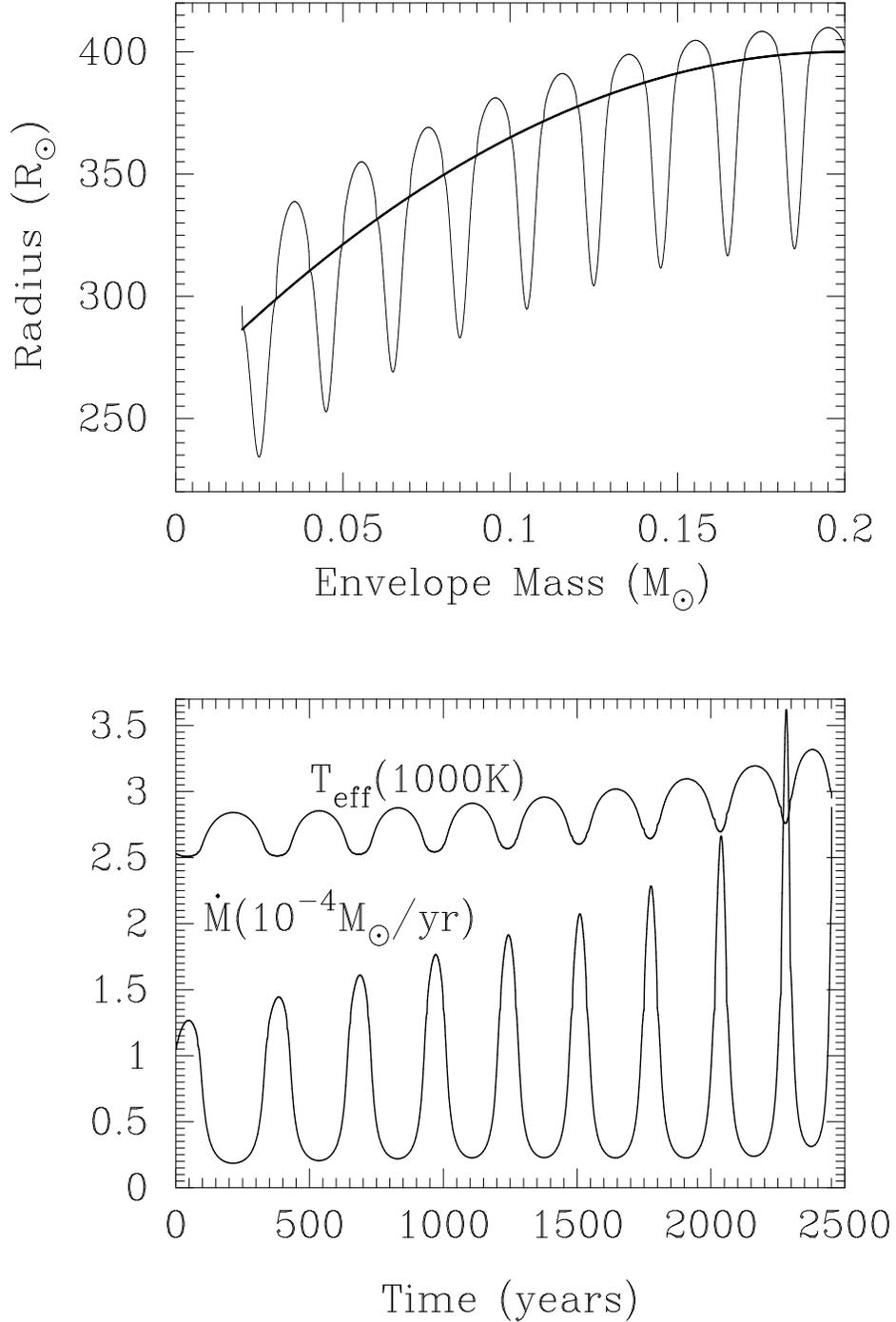}
\vskip 0.2 cm
\caption{The phenomenological upper-AGB to post-AGB evolution.
The upper panel shows the assumed variation of the envelope
radius with envelope mass.
The thick line is a crude fit to the average equilibrium radius
of several AGB evolutionary models,
while the thin line is the phenomenological postulated variation.
The lower panel shows the calculated time evolution of
the stellar effective temperature (for a stellar luminosity of
$L_{\rm eq}= 6000 L_\odot$), and the mass loss rate.
The calculation ends when the envelope mass decreases to
$0.02 M_\odot$.
Note than in the upper panel the star moves to the left during
its evolution. }
\end{figure}

To follow the evolution with time, a mass loss formula should
be given.
Again, for demonstrative purposes I use a formula that includes very
strong dependence on the temperature (or stellar radius),
as found by people studying pulsations and dust formation
(e.g., Bowen \& Willson 1991; H\"ofner \& Dorfi 1997;
Wachter et al.\ 2002), but includes dependence on
envelope mass as well, for reasons discussed in the previous
subsection (Soker \& Harpaz 2003).
The mass loss rate is taken to be
\begin{equation}
\dot M = 10^{-4} \left( \frac {R}{400 R_\odot} \right)^8
\left( \frac {M_{\rm env}}{0.2 M_\odot} \right)^{-1.5}
M_\odot \yr^{-1}
\end{equation}
Although the numerical values in this formula were chosen
to accord with the formula for $R_{\rm long}$, I suggest
that its form is more general.
With the mass loss rate as a function of envelope mass and radius
given, the equations for the evolution of the envelope
with time can be numerically integrated.
The evolution of the mass loss rate with time is given by
the lower line in the lower panel of Figure 1.
The upper line in this panel gives the effective temperature,
assuming a constant luminosity of $L_{\rm eq} = 6000 L_\odot$.
As in the rest of this section, the dynamical oscillations
(the regular Mira, etc., oscillations) on a time-scale
of $\sim 1 \yr$, are not considered.

Although very speculative, I hope that the postulated long-term
oscillations discussed here will motivate more
research into the post-AGB mass loss process.
As indicated in section 2, the very high mass loss rate,
which probably comes with a slower wind, may substantially increase
the likelihood of disk formation, hence jets, around a companion.
Finally, it is also possible that such long-term oscillations
require the envelope to be perturbed, e.g., by being spun-up
via tidal interaction with a companion.
In such a case, only systems where the companion is close, but

not too close to prevent RLOF, will have strong long-term
oscillations.

\section{Bubbles in PNs and Clusters of Galaxies}

This section summarizes the main results of three of my recent papers
(Soker 2003a,b,d).
These papers point to an interesting and nontrivial similarity in
the morphology and some non-dimensional quantities between pairs
of X-ray-deficient bubbles in clusters of galaxies and pairs
of optical-deficient bubbles in PNs.
This similarity leads me to postulate a similar formation mechanism,
hence strengthening models for PN shaping by jets (not all PNs
are shaped by jets).
The presence of dense material in the equatorial plane observed
in the two classes of bubbles constrains the jets and CFW activity
in PNs to occur while the AGB star still blows its dense wind,
or very shortly after.
Only stellar companions can account for such jets.
I then find that to inflate fat bubbles, the
opening angle of the jets must be large, i.e., the half opening
angle measured from the symmetry axis of the jets should typically
be $\alpha \gtrsim 40 ^\circ$, or the jets should precess.
For such wide-opening angle jets a collimated fast wind (CFW)
is a more appropriate term.
Narrow jets will form elongated lobes rather than fat bubbles.
I emphasize the need to include jets with a large opening angle,
i.e., $\alpha \simeq 30-70^\circ$, in simulating bubble inflation
in both PNs and clusters of galaxies (these may resemble precessing
jets, which are more difficult to simulate).

Most, or all, pairs of bubbles in clusters are found in
cooling flow clusters, i.e., those where the radiative cooling
time of the gas at their center is shorter than the age
of the cluster.
I find the term cooling flow appropriate.
Gas cools radiatively, and star formation and AGN activity
indicate that some gas is removed from the intracluster medium
(ICM). Therefore, some inflow must exist, whether of the hot,
warm, or cold medium.
The cooling rate is much below values cited a decade ago;
an appropriate term is therefore a
{\it moderate cooling flow model} (Soker et al.\ 2001).

\subsection{General arguments for jets in PNs}
The idea that jets (or CFW) shape PNs is not new. Jet shaping was
proposed by several authors to explain different morphological
features, e.g., jets (or CFW) blown by a stellar companion (Morris
1987; Soker \& Rappaport 2000) to explain bipolar PNs, and jets
blown at the final AGB phase or early post-AGB phase to form dense
blobs along the symmetry axis (Soker 1990; these blobs are termed
ansae, or FLIERs for fast low ionization emission regions), or
shape the PN (Sahai \& Trauger 1998; Sahai 2004). Basically, the
need for jet shaping came with the failure of the interacting
stellar winds (ISW) model to explain some structures observed in
PNs. (In the ISW model the shaping is due mainly to the
interaction of the fast wind blown by the central star with a
previously ejected non-spherical AGB wind; see Balick \& Frank
2002 and references therein.) This failure was pointed out already
in 1990, when I showed that the ISW can't form ansae (Soker 1990).
Later, the point-symmetric structure observed in many PNs (Sahai
\& Trauger 1998) finally killed the ISW model as the major PN
shaping process. Wind interaction still occurs and influences the
structure of PNs, but in most PNs it is not the major shaping
process. Therefore, despite the optimistic view presented by Icke
(see Rijkhorst, Icke, \& Mellema 2004) in the last APN3 meeting
when presenting his numerical simulations, I disagree with his
claim that the ISW model can account for the shaping of PNs. The
right direction to take to understand the shapes of bipolar PNs
(those with large lobes or bubbles) is to simulate AGB winds blown
simultaneously with jets (or CFW) blown by a companion
(Garcia-Arredondo \& Frank 2004), as proposed by Morris (1987) and
Soker \& Rappaport (2000). On the observational side, a search for
CFW seems to be promising (e.g., Lee, Lim, \& Kwok 2004).

Despite these arguments, and unlike the dominant view in the
clusters community, the idea of shaping by jets is controversial
in the PNs community.
I think that the similarities in the structure of pairs of bubbles
in clusters and in some PNs strongly support jets' shaping of
bubbles and lobes in PNs.
Note that not all PNs have bubbles (the more spherical cavities)
or lobes (the more elongated cavities), so not all PNs are
shaped by jets.

\subsection{The similarities}
{\it Chandra} X-ray observations of clusters of galaxies reveal the
presence of X-ray-deficient bubbles in the inner regions of many
clusters, e.g., Hydra A (McNamara et al.\ 2000),
Abell~2052, (Blanton et al.\ 2001, 2003),
A 2597 (McNamara et al.\ 2001), RBS797 (Schindler et al.\ 2001),
Abell~496 (Dupke \& White 2001), and Abell~4059 (Heinz et al.\ 2002).
These bubbles are characterized by low X-ray emissivity, implying low
density.
In most cases, the bubbles are sites of strong radio emission.
The optical morphologies of some PNs reveal
pairs of bubbles (cavities), similar in morphology to the pairs
of X-ray-deficient bubbles in clusters of galaxies.
Examples are given in table 1, and a more detailed discussion,
including more examples, are in Soker (2003a).

Despite the several orders of magnitude differences in some quantities
between clusters and PNs, the values of some non-dimensional
quantities are similar (see table 2 of Soker 2003a).
The main qualitative difference between the two classes is
the environment into which the bubbles expand.
Bubbles in clusters evolve inside the ICM, which is in
hydrostatic equilibrium; if global flow is present,
it is highly subsonic.
The bubbles in clusters move outward because of buoyancy.
In PNs the bubbles move outward as part of the global outflow
of the wind. Gravity is negligible in PNs.
However, this difference doesn't influence the
inflation phase of the bubbles.

The main relevant similarities between the two types of bubble
pairs are as follows.
(1) The most striking similarity is in the morphology.
In particular, in many cases there is a dense region in
the equatorial plane between the two bubbles,
e.g., the cluster A 2597 (McNamara et al.\ 2001)
and the Owl PN(NGC 3587: Guerrero et al.\ 2003).
(2) In some cases more than one pair of bubbles are seen,
e.g., in the Perseus cluster (Fabian et al.\ 2000, 2002) and in
the PN Hu 2-1 (Miranda et al.\ 2001b).
(3) In both types of bubbles the density inside the bubble is
2-3 orders of magnitude lower than that in the environment,
with an opposite ratio in temperatures.
(4) In both cases the typical lifetime of observed bubbles is
estimated to be $\sim 10$ times the estimated duration
of the main energy injection phase that forms the bubbles.
(5) In clusters the bubbles move subsonically, or mildly
supersonically, through the ICM.
In PNs the situation is more complicated, but this does not
much influence the inflation phase of the bubbles
(see Soker 2003a for more on these).

It is those similarities in morphologies and some non-dimensional
quantities that hint at a similar formation process in these vastly
different objects (clusters and PNs).
In clusters it is commonly accepted that pairs of bubbles are
formed by two opposite jets
(e.g., Brighenti \& Mathews 2002; Br\"uggen 2003;
Br\"uggen et al.\ 2002; Fabian et al.\ 2002;
Nulsen et al.\ 2002; Quilis, Bower, \& Balogh 2001;
Soker, Blanton, \& Sarazin 2002; Omma et al.\ 2003).
An axisymmetrical density structure of the ambient medium is
not needed to form the cluster's bubbles (only very close to the AGN,
on a scale much smaller than the bubble size, does the accretion
disk influence the flow).

The similarity in several non-dimensional quantities
suggests that if the initial flow structure is similar,
the bubble morphologies will be similar, as observed.
This leads to the following.
(1) The similar shapes strengthen the general
idea that jets (or CFW) form and shape the bubbles in PNs, as
well as other types of bipolar PNs.

(2) The low density in the bubble implies that the jets are fast,
with a speed of $> 100 \km \s^{-1}$ in PNs.
Therefore, the object launching the jets in PNs must be compact,
since the jets' speed is of the order of the escape velocity
(Livio 2000).
(3) The presence of more than one pair of bubbles in the PN Hu 2-1
 indicates, as in clusters, multiple episodic events.
(4) In clusters the surrounding density increases as radius decreases
down to the center.
The similar bubble morphologies and the presence of dense material
in the equatorial plane between the two bubbles suggests that a
similar ambient medium exists in PNs when the jets are blown.
Namely, the AGB dense wind is still active, or has ceased only
recently, when the jets are blown in PNs.
This is possible only if the jets are blown by a companion, or the
central star moves extremely rapidly from the AGB to become a compact
star that can blow fast jets.
This rapid evolution is in contrast to finding of stellar evolution
studies, and is also unlikely to explain the multiple activity
(point 3 above; Miranda et al.\ 2001b argue for a CFW that was blown
by a binary system progenitor of Hu 2-1).
One of the observational implications is that we should see evidence
of fast jets in objects that are still unambiguous AGB stars.
A good example is the system OH231.8+4.2 (Rotten Egg nebula),
for which Kastner et al.\ (1998) detect the presence of a Mira inside
this bipolar nebula which contains jets (Zijlstra et al.\ 2001).
Zijlstra et al. (2001) present evidence for jets in some OH/IR
early post-AGB stars.
There are also resolved jets near some AGB stars (e.g.,
Imai et al.\ 2002, 2003, for W34A; Hirano et al.\ 2004 and
Sahai et al.\ 2003 for  V Hydrae; Vinkovic et al.\ 2004).

\subsection{Summary: main results of the similar morphology}
The similarity in morphology and some other properties strongly supports
jets or CFW models for the shaping of pair of bubbles in PNs.
The presence of dense material in the equatorial plane constrains
the jets and CFW activity to occur while the AGB star
still blows its dense wind, or very shortly after.
The requirement that the jets and CFW be fast and the presence
of more than one pair of bubbles in, e.g., Hu 2-1,
constrains the object that blows the jets and CFW to be a compact
companion, i.e., a main sequence or a white dwarf star.
Although I considered here only PNs with well defined pairs
of closed bubbles, the results are more general in strengthening
the idea that bipolar and extreme elliptical PNs are shaped by
jets or CFW blown by an accreting companion.
Going from PNs to clusters, some determined quantities in PNs,
e.g.,  the inflating jets are non-relativistic, may help
constrain the bubble formation process in clusters.

\begin{deluxetable}{lll}
\scriptsize
\tablecaption{Similar images of PNs and clusters}
\tablehead{
\colhead{Structure} & \colhead{Clusters} & \colhead{PNs}
}
\startdata
Butterfly shape of the & Abell 478        & Roberts 22          \\
bright region; faint   & (Sun et al. 2003, & (Sahai et al. 1999,  \\
along symmetry axis    & fig 1) [1]       & fig. 1a) [2]        \\
\hline
Pairs of fat spherical & Perseus          & NGC 3587   \\
bubbles near center    & (Fabian et al.   & (Guerrero et al. 2003, \\
                       &  2000) [3]       &  fig. 1) [4]          \\
\hline
Closed bubbles         & Abell 2052            &   VV 171         \\
connected at the       & (Blanton et al. 2001,  & (Sahai 2001)   \\
equatorial plane       &  fig. 3) [5]         &   [6]          \\
\hline
Open bubbles           & M 84                  &  He 2-104           \\
connected at the       & (Finoguenov \& Jones  & (Sahai \& Trauger,  \\
equatorial plane       & 2001, fig 1) [7]       &  1998) [8]          \\
\hline
Pair of bubbles     & HCG 62             &  Hu 2-1                \\
detached from a     & (Vrtilek et al.    &  (Miranda et al. 2001b,  \\
bright center       &  2002) [9]         &  fig. 2) [10]          \\
\hline
Point-symmetric     & Hydra A              & NGC 6537          \\
elongated lobes     & (McNamara et al.     & (Balick 2000,    \\
                    &  2000, fig. 1) [11]  & fig. 2) [12] \\
\hline
Pairs of bright     &  Cygnus A             &  NGC 7009       \\
bullets along the   &  (Smith et al.        &  (Balick et al.         \\
symmetry axis       &   2002, fig. 1) [13]  &   1998, fig. 1,4) [14]  \\
\enddata
\tablecomments{
Similar images of bubbles in clusters of galaxies and
planetary nebulae (PNs).
In clusters these are X-ray images (e.g., with X-ray deficient bubbles),
while in PNs they are optical images (e.g., with optical
deficient bubbles).
In the first five pairs of images the similarity is of high
degree. In the last two pairs of images the similarity
between the cluster and the PN is less. }
\end{deluxetable}

\noindent {\bf Comments to table 1:}
The images of all these objects are summarized in a {\it powerpoint}
file I presented during the APN3 meeting (2003). The site:
\newline
http://www.astro.washington.edu/balick/APN/APN\_talks\_posters.html
\newline
(go the `ppt' file in the ``discussion'' of session 13).

Free access to individual images are at these sites:
\newline
[1] http://arxiv.org/PS\_cache/astro-ph/pdf/0210/0210054.pdf
\newline
[2] http://ad.usno.navy.mil/pne/images/rob22.jpg
(Terzian \& Hajian 2000)
\newline
[3] http://arxiv.org/PS\_cache/astro-ph/pdf/0007/0007456.pdf
\newline
[4] http://arxiv.org/PS\_cache/astro-ph/pdf/0303/0303056.pdf
\newline
[5] http://arxiv.org/PS\_cache/astro-ph/pdf/0107/0107221.pdf
\newline
[6] http://ad.usno.navy.mil/pne/images/vv171.jpg
\newline
[7] http://arxiv.org/PS\_cache/astro-ph/pdf/0010/0010450.pdf
\newline
[8] http://ad.usno.navy.mil/pne/images/he2\_104.jpg
\newline
[9] http://chandra.harvard.edu/photo/cycle1/hcg62/index.html
\newline
[10a] http://arxiv.org/PS\_cache/astro-ph/pdf/0009/0009396.pdf
\newline
also: [10b] http://ad.usno.navy.mil/pne/images/hu21\_ha.gif
\newline
[11] http://arxiv.org/PS\_cache/astro-ph/pdf/0001/0001402.pdf
\newline
[12] http://ad.usno.navy.mil/pne/images/ngc6537.jpg
\newline
[13] http://arxiv.org/PS\_cache/astro-ph/ps/0109/0109488.f1.gif
\newline
[14a] http://ad.usno.navy.mil/pne/images/ngc7009.jpg
\newline
see also (Goncalves et al.\ 2003, fig. 1)
\newline
[14b] http://arxiv.org/PS\_cache/astro-ph/pdf/0307/0307265.pdf

\newpage

\section{Relevant Related Objects}

The main point of this review, like that of many of my papers in the
last 15 years, is that binary interaction, with a stellar
or a substellar companion (planets or brown dwarfs), can
account for the shaping of PNs.
PN research can contribute a lot to the study of shaping
of some other astrophysical systems, and at the same time benefit
from research conducted to understand these and other objects.
In this section I list some of these objects.

\noindent {\bf Symbiotic Systems.}
Schwarz gave a talk (Schwarz \& Monteiro 2004) and Corradi (2004)
led a discussion on the relation of symbiotic systems to bipolar
PNs$-$those with lobes (or fat-bubbles) and an equatorial waist
between them.
Although the relation of PNs to symbiotic systems was noted by
several people in the past, Corradi and Schwarz (1995,
and other papers) led the research along this line.
The morphological similarity between some symbiotic nebulae
and some bipolar PNs strongly suggests that bipolar PNs are
formed by binary systems, most with the accreting companion
outside the AGB envelope.
The accumulating evidances for jets in symbiotic systems
(Kellogg, Pedelty, \& Lyon 2001; Brocksopp et al.\ 2003)
add to their connection to PNs and some other objects
listed here.
For example, note the similarity in the X-ray jet in R Aqr
(Kellog et al.\ 2001) and the X-ray jet in Mz 3
(Kastner et al.\ 2003).
Other arguments for binary progenitor models of bipolar PNs,
such as the class of post-AGB stars with a companion and a
circumbinary disk (Jura 2004; Van Winckel 2003, 2004),
are summarized in Soker (1998).

\noindent{\bf Supersoft X-ray Sources.}
Supersoft X-ray sources are thought to be white dwarfs accreting at
rates of $10^{-8}-10^{-7} M_\odot \yr^{-1}$ from a companion, and
sustaining nuclear burning on their surface (e.g., Greiner 1996).
Fast, $\sim 1000-5000 \km \s^{-1}$, collimated outflows have been
observed in some supersoft X-ray sources
(Southwell, Livio, \& Pringle 1997; Becker et al. 1998; Motch 1998).
These systems teach us that accreting WDs can blow jets
(Soker \& Rappaport 2000).
Bipolar PNs hint that bipolar nebulae should exist around some
supersoft X-ray sources, even if very faint.

\noindent{\bf Novae.} The relation of novae to PNs was reviewed
by Bode (2004) at the APN3 meeting.
After the novae eruption, the system enters a short common envelope
phase. The axisymmetrical structure of many novae eruptions shows
that common envelope phase can lead to axisymmetrical structures.
About 20 PNs are known, and many more are expected, to
harbor close binary systems (Bond 2000; Sorensen \& Pollacco 2004;
Hillwig 2004; De Marco et al.\ 2004)
which went through a common envelope phase.

\noindent{\bf R Coronae Borealis stars.}
These are carbon-rich hydrogen-poor stars, with effective temperatures
more than twice those of AGB stars. They are known to form dust
sporadically on local spots very close to the surface
(at APN3 these objects were discussed by Clayton 2004).
Their relevance is that they show that dust can be easily formed
very close to the surface of giant stars, in a non-spherical
configuration (Soker \& Clayton 1999).

\noindent{\bf Young stellar objects.}
Many YSO are known to blow jets. Similar jets in PNs
(Lee \& Sahai 2004), e.g., similar outflow speeds, strengthen the
case for an accretion disk as the source of the jets in PNs, and most
likely indicate that the companion blowing the jet is a main-sequence
companion.
(Much faster jets are likely to originate at an accreting white dwarf,
as in supersoft X-ray sources.)

\noindent{\bf Massive stars: WR 98A.} This massive binary system,
where one of the components is a massive WR star, has a circumbinary
matter in a spiral structure (Monnier, Tuthill, \& Danchi 1999).
This spiral structure results from a binary interaction.
Spiral structure is expected to be formed by some binary-systems
progenitors of PNs (Soker 1994; Mastrodemos \& Morris 1999).
The spiral is expected to be smeared quite fast, in particular
after ionization starts.
WR 98A and similar systems suggest that careful observations
of proto-PNs, possibly by dividing two images at different
spectral lines, may reveal spiral structure.

\noindent{\bf Massive stars: $\eta$ Car.}
Gehrz (2004) reviewed $\eta$ Car at the APN3 meeting.
This system which is known to harbor a binary system, shows that
a binary system is behind the bipolar structure.
The arguments given here, i.e., that the companion blows
jets that form the lobes of bipolar PNs, suggest that
in $\eta$ Car as well the bipolar structure was formed
by jets blown by the accreting companion (Soker 2001b)

\noindent{\bf Massive stars: SN 1987A.}
Sugerman \& Crotts (2004) review the structure of the material around
the progenitor of SN 1987A.
The basic structure is similar to that of some bipolar PNs.
This suggests, as was noted in the past, that a binary
companion most likely shaped the circumstellar matter around
SN 1987A, and that it accreted mass and blew jets.

\noindent{\bf Massive stars: $\rho$ Cassiopeiae.}
This yellow hypergiant stars is known to go through non-periodic
outbursts (Lobel et al.\ 2003).
The outbursts are suggested to result from envelope instabilities
(Lobel et al.\ 2003).
This may be related to the postulated long-term oscillations of
upper AGB stars, in showing that envelopes of giant stars may be
unstable on time-scales of $10-1000 \yr$, i.e., longer than the
dynamical time.

\noindent{\bf Clusters of galaxies.}
These objects are different in nature from all other related
objects listed above, as these are not stellar objects,
and the length-scales involved are $\sim 5$ orders of magnitude larger
than those involved in PNs.
However, the similar structure of bubbles in clusters and bubbles
in PNs is striking and nontrivial.
I argue that the similarity in structure and several non-dimensional
quantities (Soker 2003a) has implications for the formation mechanisms
of fat-bubbles in PNs:
they are formed by jets blown by a companion while the progenitor
is still an AGB star, or a very young post-AGB star.
In Soker (2003b) I argue that some of the fat-bubbles in clusters
and PNs are likely to be formed by jets having a wide
opening angle (collimated fast wind; CFW), or precessing jets.
Rayleigh-Taylor instability at shell's front is discussed in
Soker (2003d).

\noindent{\bf Subdwarf B binaries.}
These objects are related to the question of the transition
from spherical to axisymmetrical mass loss geometry just as the
star is about to leave the AGB. In section 2 I explain this in
the paradigm of stellar binary interaction.
The subdwarf B binaries (sdB) composed of a helium core of mass
$\sim 0.5 M_\odot$ and a very thin hydrogen layer of $< 0.02 M_\odot$,
and a companion, either a main sequence star of a white dwarf,
or even a brown dwarf companion (Rauch 2003),
with orbital periods in the range $\sim 1$~hour$-$1000~days
(see Morales-Rueda, Maxted, \& Marsh 2003, and references therein).
It is not clear how these stars (the helium core stars) lose most of
their hydrogen-rich envelope on the red giant branch (RGB) but
still managed to ignite helium (Morales-Rueda et al.\ 2003).
It is probably that the binary interaction that causes most
of the envelope to be lost occurs when these stars are on the
upper RGB, and just about to ignite helium.
The rapid interaction with stars on the upper RGB can be
understood from the evolution of the stellar radius with core mass.
The analytical approximate relations from Iben \& Tutukov (1984)
for RGB stars which are descendant of population I main sequence
stars in the mass range $0.8 < M/M_\odot < 2.2$ read
\begin{equation}
\frac {R_g}{R_\odot} =10^{3.5}
\left( \frac {M_c}{M_\odot} \right)^4 \qquad {\rm and} \qquad
\dot M_c  =10^{-5.36}
\left( \frac {M_c}{M_\odot} \right)^{6.6} M_\odot \yr^{-1},
\end{equation}
where $M_c$ is the core mass, and $R_g$ the RGB radius.
These relations show that when the RGB star reaches large radii,
its core mass is already large, and the evolution becomes
more rapid as the core mass and radius increase.
Considering that tidal interaction evolves on short time
scales as well (section 2), it is expected that many RGB stars
will interact with their binary companion during a short time
when they are large and their core is massive.
(These stars can then ignite helium in the core while cooling
after envelope collapse; D'Cruz et al.\ 1996.)
Similar rapid evolution of binary interaction occurs on the
upper AGB.
This is the relation of these binary stars to PNs and proto-PNs
showing rapid transition from spherical to axisymmetrical mass loss.

\acknowledgments
This review was initiated by two meetings I attended in the summer of
2003: The Riddle of Cooling Flows in Galaxies and Clusters of Galaxies,
and Asymmetrical Planetary Nebulae III.
I benefited from discussions with many people during these two
meetings. Among these are
James Binney, Adam Frankowski, Joel Kastner, Raghvendra Sahai,
Hans Van Winckel, Jan Vrtilek, Ray White, and Albert Zijlstra.
Special thanks to my long-time collaborator Amos Harpaz,
for making possible the study of the AGB and post-AGB stellar
evolution, and for many useful discussions.
This research was supported in part by the
Israel Science Foundation.


\end{document}